\documentclass[final,1p,times]{elsarticle}
\usepackage{amssymb}
\journal{Nuclear Physics A}

\begin{document}

\begin{frontmatter}



\title{Strange mesons from SIS to FAIR}


\author{L. Tolos$^1$, D. Cabrera$^2$, A. Polls$^3$ and A. Ramos$^3$}

\address{$^1$Theory Group. KVI. University of Groningen, \\
Zernikelaan 25, 9747 AA Groningen, The Netherlands}

\address{$^2$Departamento de F\'{\i}sica Te\'orica II, Universidad Complutense,\\
28040 Madrid, Spain}

\address{$^3$Departament d'Estructura i Constituents de la Mat\`eria,\\
Universitat de Barcelona,
Diagonal 647, 08028 Barcelona, Spain}

\begin{abstract}
The properties of $K$ and $\bar K$ mesons in nuclear matter at finite
temperature are obtained from a chiral unitary approach in coupled channels which incorporates the $s$- and $p$-waves of the kaon-nucleon interaction. The
in-medium solution accounts for Pauli blocking effects, mean-field binding on
all the baryons involved, and $\pi$ and kaon self-energies.  
The $\bar K$ spectral function spreads over a wide range of energies, 
reflecting the melting of the $\Lambda (1405)$ resonance and the contribution
of hyperon-hole components at finite temperature. In the $KN$ sector, the quasi-particle peak is considerably broadened with increasing density and temperature.
We also study the energy weighted sum rules of the kaon propagator by matching the Dyson form of the
propagator with its spectral Lehmann representation at low and high 
energies. The sum rules for the lower energy weights are fulfilled
satisfactorily and reflect the contributions from the different 
quasi-particle and collective modes of the spectral function. We analyze the sensitivity of the
sum rules to the distribution of spectral strength and their 
usefulness as quality tests of model calculations. 
\end{abstract}


\end{frontmatter}



\section{Introduction}
\label{introduction}
The properties of strange hadrons in hot and dense matter is a matter of extensive study due to the implications for the phenomenology of exotic atoms \cite{Friedman:2007zz} as well as heavy-ion collisions from SIS \cite{Fuchs:2005zg} to FAIR \cite{fair} energies at GSI . Whereas the interaction of $\bar K N$ is repulsive at threshold, the phenomenology of antiskaonic atoms shows that the $\bar K$ feels an attractive potential at low densities. This attraction is a consequence of the modified $s$-wave $\Lambda(1405)$ resonance in the medium due to Pauli blocking effects \cite{Koch} together with the self-consistent consideration of the $\bar K$ self-energy \cite{Lutz} and the inclusion of self-energies of the mesons and baryons in the intermediate states \cite{Ramos:1999ku}. Attraction of the order of -50 MeV at normal nuclear matter density, $\rho_0=0.17 \,{\rm fm^{-3}}$, is obtained by different approaches, such as unitarizated theories in coupled channels based on chiral dynamics \cite{Ramos:1999ku} and meson-exchange models \cite{Tolos01,Tolos02}. In fact, recent few-body calculations \cite{shevchenko,Ikeda:2007nz,dote_hyp06} predict few-nucleon kaonic states bound only by 50--80 MeV and having large widths of the order on 100 MeV, thereby disclaiming the finding of deeply kaonic bound states \cite{akaishi}. Moreover, the knowledge of higher-partial waves beyond s-wave \cite{Tolos:2006ny,Lutz:2007bh,Tolos:2008di} becomes essential for relativistic heavy-ion experiments at beam energies below 2AGeV \cite{Fuchs:2005zg}.

In order to test the quality of the calculation of the kaon self-energy, we can exploit the analytical properties of the kaon single-particle propagator, which  impose some
constraints on both the many-body formalism and the interaction model.  An excellent tool to analyze these 
constraints is provided by the energy-weighted sum rules (EWSRs) of the
single-particle spectral functions. Therefore, in this paper we review the properties of the strange ($K$ and $\bar K$) mesons in hot and dense matter and  we study the first EWSRs of the
kaon single-particle spectral functions.

\section{Strange mesons in nuclear matter}

The  kaon self-energies in symmetric nuclear matter at finite
temperature are obtained from the in-medium kaon-nucleon interaction within a chiral unitary approach. The model incorporates the 
$s$- and $p$-waves of the kaon-nucleon interaction \cite{Tolos:2008di}. 

The $s$-wave amplitude arises from the Weinberg-Tomozawa term of the chiral Lagrangian. Unitarization in coupled channels is imposed by solving the Bethe-Salpeter equation with on-shell amplitudes ($T$) and a cutoff regularization. The unitarized $\bar K N$ amplitude generates dynamically the $\Lambda(1405)$
resonance in the $I=0$ channel and provides a satisfactory description of
low-energy scattering observables. The $K N$ effective interaction is also obtained using the Bethe-Salpeter with the same cutoff regulator.  The in-medium solution of the $s$-wave amplitude accounts for Pauli-blocking
effects, mean-field binding on the nucleons and hyperons via a $\sigma-\omega$
model, and the dressing of the pion and kaon propagators. The self-energy is then obtained in a self-consistent manner summing the transition amplitude $T$ for the different isospins over the nucleon Fermi distribution at a given temperature, $n(\vec{q},T)$, as 
\begin{eqnarray}
\Pi(q_0,{\vec q},T)= \int \frac{d^3p}{(2\pi)^3}\, n(\vec{p},T) \,
[\, {T}^{(I=0)} (P_0,\vec{P},T) +
3 \, {T}^{(I=1)} (P_0,\vec{P},T)\, ]\ , \label{eq:selfd}
\end{eqnarray}
where $P_0=q_0+E_N(\vec{p},T)$ and $\vec{P}=\vec{q}+\vec{p}$ are
the total energy and momentum of the kaon-nucleon pair in the nuclear
matter rest frame, and ($q_0$,$\vec{q}\,$) and ($E_N$,$\vec{p}$\,) stand  for
the energy and momentum of the kaon and nucleon, respectively, also in this
frame. In the case of $\bar K$ meson the model also includes, in addition, a $p$-wave contribution to the self-energy from hyperon-hole ($Yh$) excitations, where $Y$ stands for $\Lambda$, $\Sigma$ and
$\Sigma^*$ components. For the $K$ meson the $p$-wave self-energy results from $YN^{-1}$ excitations in crossed kinematics. The spectral function depicted in the following results from the imaginary part of the in-medium kaon propagator.

%%%%%%%%%%%%%%%%%%%%%%%%%%%%%%%%%%%%%%%%%%%%%%%%%%%
\begin{figure}[htb]
\begin{center}
\includegraphics[height=6 cm, width=6.6 cm]{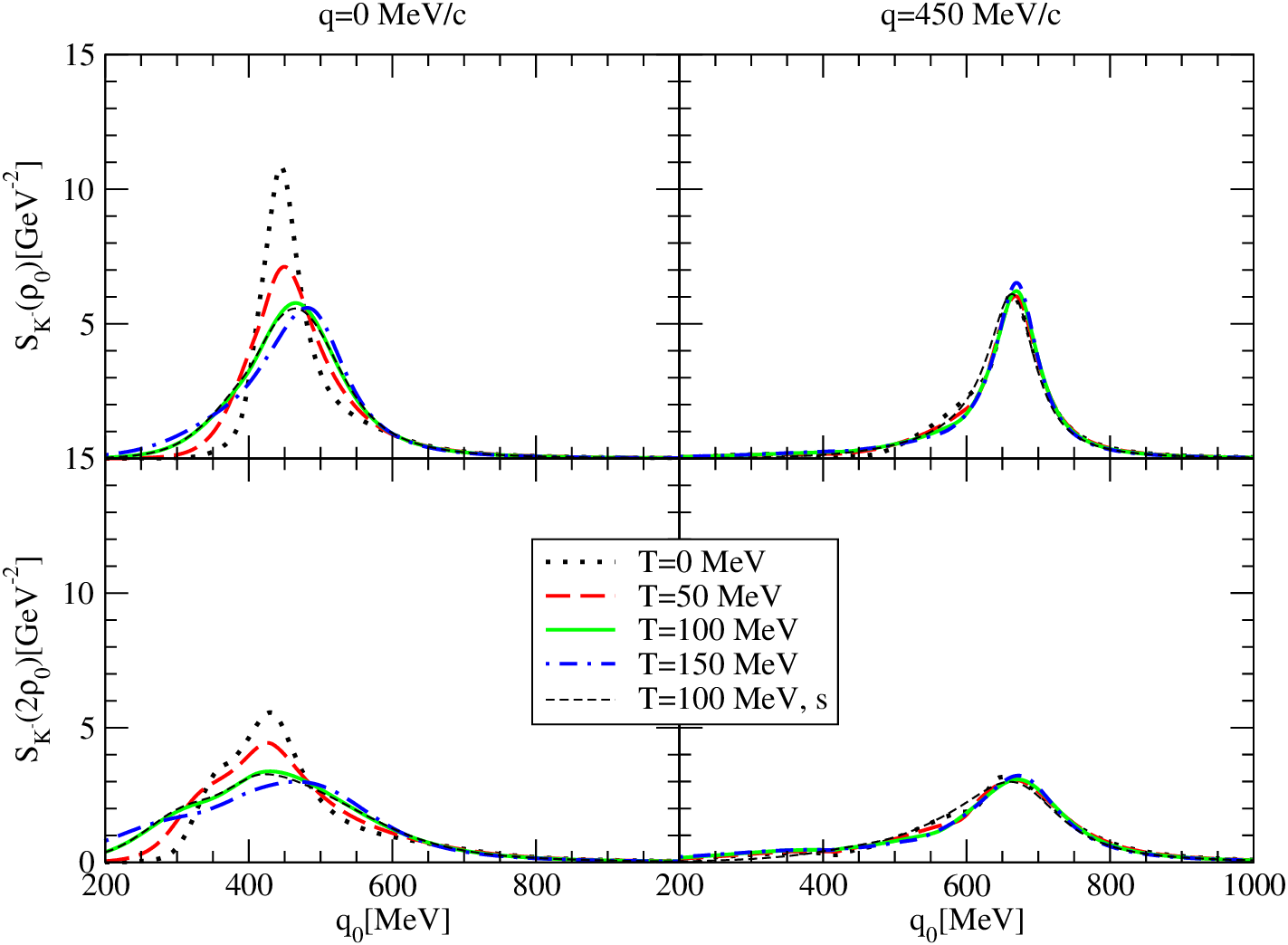}
\hfill
\includegraphics[height=6 cm, width=6.6 cm]{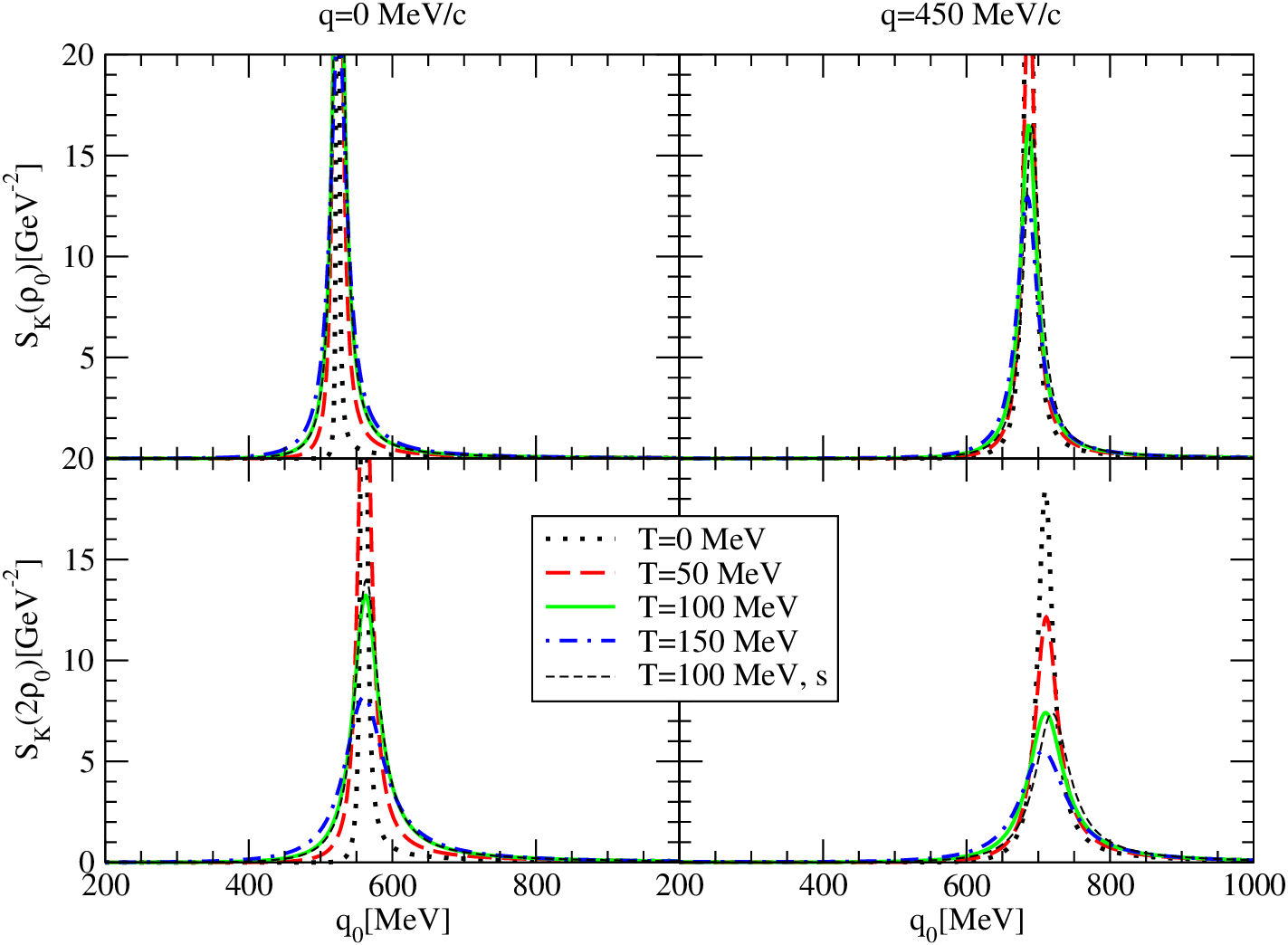}
\caption{ Evolution of the $\bar K$ and $K$ spectral functions with density and temperature for two momenta.}
 \label{fig1}
\end{center}
\end{figure}
%%%%%%%%%%%%%%%%%%%%%%%%%%%%%%%%%%%%%%%%%%%%%%%%%%%%

The evolution with density and temperature of the $\bar{K}$ and $K$ spectral functions are depicted in Fig.~\ref{fig1}. The $\bar K$ spectral function (left plot) shows a strong mixing between the quasi-particle peak and the $\Lambda(1405)N^{-1}$ and  $Y(=\Lambda, \Sigma , \Sigma^*)N^{-1}$ excitations. The effect of these $p$-wave $YN^{-1}$ subthreshold excitations is
repulsive for the $\bar K$ potential, compensating in part the attraction
from the $s$-wave ${\bar K} N$ interaction.  Temperature softens the
$p$-wave contributions to the spectral function at the quasi-particle energy. Moreover, together with the $s$-wave mechanisms, the $p$-wave self-energy
provides a low-energy tail which spreads the spectral function considerably. Increasing the density  dilutes the spectral function even further. As for the $K$ spectral function (right plot), the $K$ meson is described by a narrow quasi-particle peak which dilutes 
with temperature and density as the phase space for 
$KN$ states increases. The $s$-wave repulsive
self-energy  translates into a shift of the $K$ spectral function to
higher energies with increasing density. In contrast to the $\bar K N$ case,
the inclusion of $p$-waves has a mild attractive effect on the kaon
self-energy (compare thin-dashed lines to solid lines at $T=100$ MeV and $q=450$
MeV/c).

%%%%%%%%%%%%%%%%%%%%%%%%%%%%%%%
\begin{figure}[htb]
\begin{center}
\includegraphics[height=6.2 cm, width=6 cm]{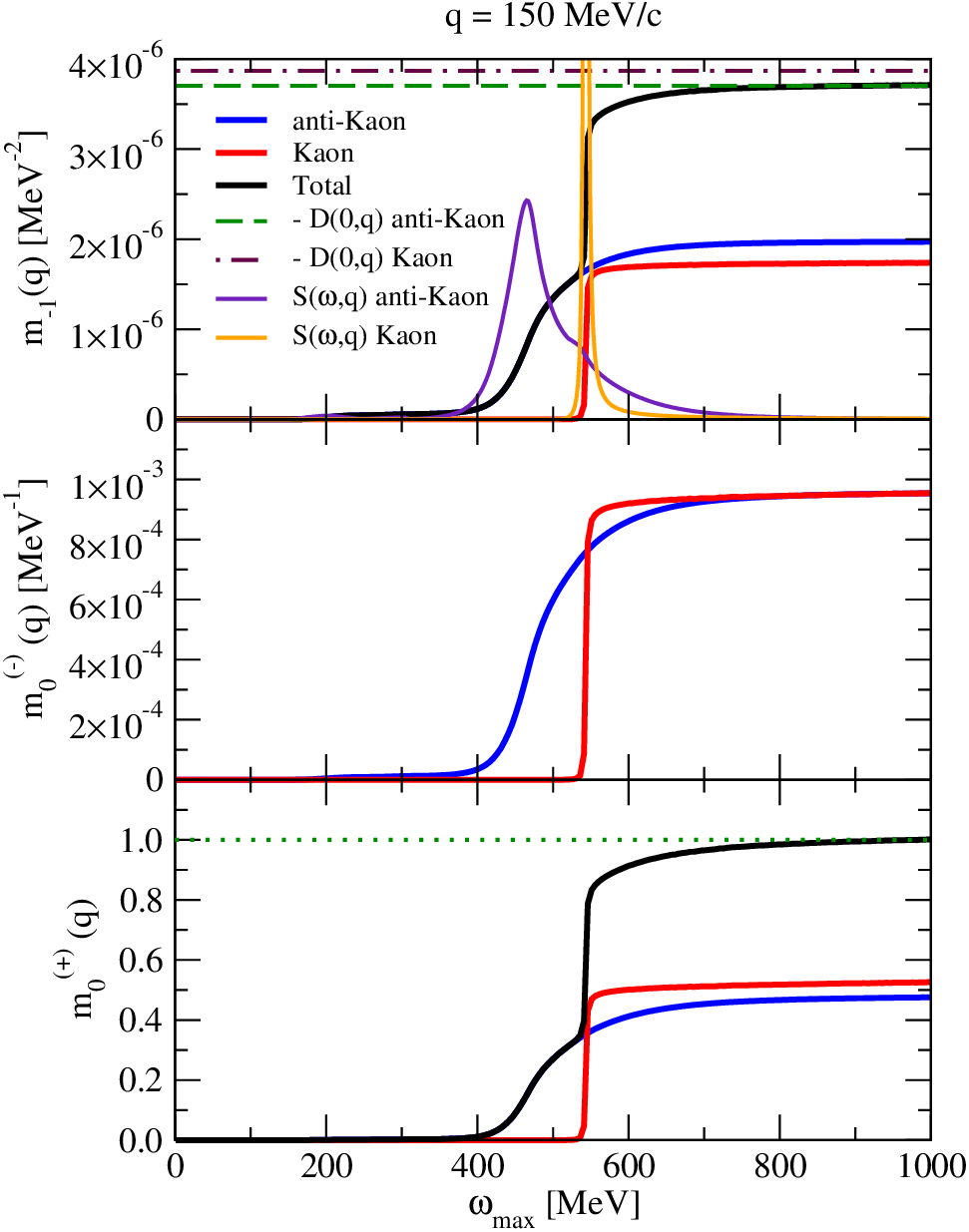}
%\includegraphics[height=6 cm, width=6 cm]{Sum-rule-KKbar-minusone-r0-T1-variousq-v2.eps}
%\hfill
%\includegraphics[height=5.8 cm, width=6 cm]{Sum-rule-KKbar-minusone-r0-T100-variousq-v2.eps}
\caption{ $m_{-1}$, $m_{0}^{(-)}$ and $m_{0}^{(+)}$ sum rules for the $K$ and $\bar K$ spectral functions at
$q=150$~MeV/c, $\rho=\rho_0$ and $T=0$ MeV. The $\bar K$ and $K$ spectral
functions are also displayed for reference.}
 \label{fig2}
\end{center}
\end{figure}

%%%%%%%%%%%%%%%%%%%%%%%%%%%%%%%%%%%%%5

\section{Energy weighted sum rules for kaons}

The EWSRs are obtained from matching the Dyson form of the meson propagator with its spectral Lehmann representation at low and high 
energies \cite{ewsr}. The first EWSRs in the high-energy limit expansion, $m_0^{(\mp)}$, together with the zero energy EWSR, $m_{-1}$, are given by
\begin{eqnarray}
m_{-1}&:&
\int_0^{\infty} \textrm{d}\omega \,
\frac{1}{\omega} \, [ S_{\bar K}(\omega,\vec{q}\,;\rho,T) + S_{K}(\omega,\vec{q}\,;\rho,T)]
=
\frac{1}{\omega_{\bar K}^2(\vec{q}\,) + \Pi_{\bar K}(0,\vec{q}\,;\rho,T)} \ ,
\\
m_{0}^{(\mp)}&:&  
\int_0^{\infty} \textrm{d}\omega \,
[ S_{\bar K}(\omega,\vec{q}\;\rho,T) - S_{K}(\omega,\vec{q}\,;\rho,T) ] = 0
\nonumber \\
& & 
\int_0^{\infty} \textrm{d}\omega \,
\omega \, [ S_{\bar K}(\omega,\vec{q}\,;\rho,T) + S_{K}(\omega,\vec{q}\,;\rho, T) ] = 1 \ .
%\nonumber \\
%\\
%m_{1}^{(\mp)}&:&  
%\int_0^{\infty} \textrm{d}\omega \,
%\omega^2 \, [ S_{\bar K}(\omega,\vec{q}\,;\rho,T) - S_{K}(\omega,\vec{q}\,;\rho,T) ] = 0
%\nonumber \\
%& & \ \ \ 
%\int_0^{\infty} \textrm{d}\omega \,
%\omega^3 \, [ S_{\bar K}(\omega,\vec{q}\,;\rho,T) + S_{K}(\omega,\vec{q}\,;\rho,T) ] 
%=
%\omega_{K}^2(\vec{q}\,) + \Pi_{\bar K}^{\infty}(\vec{q}\,;\rho,T) \ .
%\nonumber \\
\end{eqnarray}

The sum rules for the antikaon propagator are shown in Fig.~\ref{fig2} as a function of the upper integral limit for
$\rho=\rho_0$, $T=0 \ {\rm MeV}$ and $q=150~$MeV/c. The contributions from $\bar K$ and $K$ to the l.h.s. of the sum rule are depicted separately. The $\bar K$ and $K$ spectral functions are also shown for reference in arbitrary units. Note that saturation is progressively shifted to higher energies as we examine sum rules involving higher order weights in energy.

The l.h.s. of the $m_{-1}$ sum rule (upper panel) saturates a few hundred MeV beyond the quasiparticle peak, following the behaviour of the $\bar K$ and $K$ spectral functions. We have also plotted the r.h.s. of the $m_{-1}$ sum
rule both for the antikaon and kaon, namely their off-shell propagators
evaluated at zero energy (modulo a minus sign). The difference between both values reflects the violation of crossing symmetry present in the chiral model for the kaon and antikaon self-energies as we neglect the explicit
$t$-channel exchange of a meson-baryon pair. However, we  may still expect
the saturated value of the l.h.s. of the $m_{-1}$ sum-rule to provide a
constraint for the value of the zero-mode propagator appearing on the r.h.s, because the most of the strength sets in at energies of the order of the meson
mass, where the  neglected  terms of the $K (\bar K)N$ amplitudes are
irrelevant. The $m_0^{(-)}$ sum rule shows that the areas subtended by the $K$ and $\bar K$ spectral functions coincide (middle panel). The fullfilment of this sum rule is, however, far from trivial because, although the $\bar K$ and $K$ spectral
functions are related by the retardation property, $S_{\bar
K}(-\omega)=-S_K(\omega)$, the actual calculation of the meson self-energies is
done exclusively for positive meson energies. And, finally, the $m_0^{(+)}$ sum rule (lower panel) saturates to one independently of the meson momentum, nuclear density or temperature, thus posing a strong constraint on the accuracy of the calculations. 

Those sum rules have been also tested satisfactorily for higher momenta and temperature \cite{ewsr}.
As the meson momentum is increased, the saturation of the integral part of the sum rules is progressively shifted to higher energies, following the strength of the spectral distribution. At finite temperature the dilution of the $\bar K$ and $K$ spectral functions with increasing thermal phase space contributes substantially to the l.h.s. of the sum rule below
the quasi-particle peak.



\end{document}